\documentclass[aip, pop, amsmath, amssymb, floatfix, reprint]{revtex4-1}
\usepackage{subcaption}
\usepackage{amsfonts} 
\usepackage{amsmath} 
\usepackage{amssymb}
\usepackage{physics}
\usepackage{graphicx, nicefrac}
\usepackage{float}
\usepackage{blindtext}
\usepackage{lineno}
\usepackage{cancel}
\usepackage{scalerel}
\usepackage{float}
\usepackage[normalem]{ulem}
\usepackage{xfrac}
\usepackage{bm}
\usepackage[usenames, dvipsnames]{color} 
\usepackage{dcolumn}
\usepackage{epic} 
\usepackage{epsfig}
\usepackage{eucal} 
\usepackage{esdiff}
\usepackage{graphicx}
\usepackage[breaklinks,colorlinks = true,linkcolor = magenta,urlcolor=magenta,citecolor=red]{hyperref}
\usepackage{mathrsfs}
\usepackage{soul}
\usepackage{textcomp}
\usepackage{times}
\usepackage{verbatim}
\usepackage{wrapfig} 
\usepackage{xy} 
\usepackage{lineno}
\usepackage{import}
\newcommand{\ang}[1]{\ensuremath{\left\langle {#1} \right\rangle}}

\begin{document}
\title{Lagrangian intermittency and vertical confinement in stably stratified turbulence}
\author{Arun Kumar Varanasi}
\email{arun.target@gmail.com}
\affiliation{International Centre for Theoretical Sciences, Tata Institute of Fundamental Research, Bengaluru 560089, India}
\begin{abstract} 

We investigate the Lagrangian properties of homogeneous, stratified
	turbulence at different Brunt-V\"ais\"al\"a frequencies.
	We show increasing vertical confinement of trajectories with increasing
	stratification strength highlighting the predominantly horizontal
	dynamics of these flows. We characterize the intermittent properties of
	the fluid motion by computing the probability distributions of the
	horizontal and vertical Lagrangian velocity increments and Lagrangian
	horizontal velocity structure functions, and find that horizontal
	velocity differences display intermittency at higher orders. Our results provide an understanding
	of the Lagrangian dynamics in stratified turbulence, and highlight the
	importance of buoyancy forces in shaping the Lagrangian statistics in
	these flows.

\end{abstract}
\date{}
\maketitle

An understanding of the small-scale physics of the ocean and atmosphere is
closely tied to our understanding of mixing and transport processes in these
planetary systems \cite{thorpe2005,ferrari2009ocean}.  These issues typically
reduce to important questions on the dynamics of \textit{suspended particles}
in such flows. For example, the core of a cloud can be modeled reasonably as
\textit{just} a (dilute) suspension of ice crystals and water droplets. A
drastic but useful simplification is to approximate the carrier flow using
theoretically tractable simple models. Examples of such flows include shear,
simple vortical, laminar, and smooth but chaotic flows. However both oceanic
and atmospheric flows are notoriously turbulent. By construction simpler model
flows share the same \textit{smoothness} of turbulence at the particle scale.
What they lack however is a key feature of turbulence namely intermittency
\cite{Frisch-CUP,frisch1980fully}.  The intermittent nature of homogeneous and
isotropic turbulence shows itself, for example,  in the large tails of the
distributions of spatial (Eulerian) and temporal (Lagrangian) velocity
increments as well as velocity gradients. While a theory for intermittency
starting with the incompressible Navier-Stokes equation remains elusive, the
evidence that flows are intermittent have been documented in numerical
simulations~\cite{siggia1981numerical,ray2018non,pandit2009statistical},
observations~\cite{van1970structure} and
experiments~\cite{batchelor1949nature}. This deviation from Gaussian
distributions has a striking manifestation in the strong deviation (from
theoretical estimates) of the scaling exponents associated with the Eulerian
and Lagrangian structure functions of velocity increments
\cite{benzi2010inertial}.  Given the complexity of turbulence, it is legitimate
to ask whether \textit{turbulence} matters when investigating the small-scale
physics of mixing and transport.  Recent studies suggest that the answer is
yes. For example, the rate at which water droplets in a warm cloud coalesce and
hence grow to sizes which precipitate is a direct outcome of the flow being
turbulent and hence
intermittent~\cite{bec2016abrupt,bec2014gravity,devenish2012droplet,falkovich2002acceleration}.
These are important results and underline the fact that turbulence cannot be
neglected to understand such phenomena in the atmospheric, oceanic and
geophysical contexts. Furthermore, while fluid intermittency remains one of the
outstanding questions of turbulence, it is only recently that we have begun to
fully appreciate its leading order contribution to problems which have a more
direct bearing beyond the somewhat more conceptual challenges of mathematics or
statistical physics. 

Homogeneity, isotropy and constant density are of course (often justifiable at
small scales) simplifications: Real atmospheric and oceanic flows are density
stratified. While there have been several studies of stratified turbulence
\cite{riley1981direct,feraco2018vertical,rorai2014turbulence,rorai2015stably,herring1989numerical,kimura2012energy,colm2020open},
especially in the marine context with the allied, fundamental questions of
carbon sequestration and marine snow, for example, the intermittent nature of
homogeneous, stratified --- and hence anisotropic --- flows have received far
less attention than what it perhaps deserves. As we discussed above, work on
transport and mixing in homogeneous, isotropic and constant density
incompressible turbulence shows that a proper understanding of phenomena such
as collisions, coalescences, and gravitational settling of sediments --- just
as important in stratified turbulence --- must factor in the intermittent
nature of the carrier flow. Hence, in this paper we undertake the characterisation of the Lagrangian intermittency of stratified
turbulence and leave the question of how it affects, for example, the
concentration field of microorganisms and marine snow, for future work. 

As is clear from the discussion above, for problems involving stratified media
the flow needs to be modeled beyond the constant density Navier-Stokes
equation\cite{varanasi2022motion}. Typically, density stratification is chosen to be linear but when
such flows are driven (by an external force) to a turbulent state, fluctuations
in the density field ($\rho$) naturally arise. Hence, by assuming a Cartesian
coordinate system where the gravity $-g\bf{\hat{z}}$ acts downward along the
vertical $\bf{\hat{z}}$ axis, a density profile $\rho=\rho_0-\gamma z+\rho_f$
mimics a stratified turbulent fluid; in the absence of turbulence, $\rho_f =
0$, leading to a stable stratification where the density decreases linearly
with height, at a rate $\gamma$, along the vertical direction. 

Momentum conservation ensures that the incompressible velocity field ${\bf u}$ satisfies
\begin{equation}
	\rho\left [\pdv{\vb{u}}{t}+\vb{u}\cdot\nabla\vb{u}\right ]=-\nabla p_0 + \mu \nabla^2\vb{u}-\rho g {\bf \hat{z}}+\vb{F}
\label{u-general}
\end{equation}
where $p_0$ is the pressure field, $\mu$ is the bulk viscosity and $\vb{F}$ is
an external drive. In the ideal, inviscid $\mu = 0$, unforced ${\bf F} = 0$
limit, any displacement of a parcel of fluid will lead to its oscillations with
a natural, Brunt-V\"ais\"al\"a  frequency $N = \sqrt{\frac{g\gamma}{\rho_0}}$.
Further, by taking the Boussinesq approximation of replacing $\rho$ with
$\rho_0$ in the right hand side of Eq.~\eqref{u-general}, and setting $p = p_0
+ \rho_0 gz - \frac{\gamma gz^2}{2}$,  we obtain (with the forcing term ${\bf f}$ 
normalised by the density)

\begin{equation}
	\pdv{\vb{u}}{t}+\vb{u}\cdot\nabla\vb{u}=-\frac{\nabla p}{\rho_0} + \nu \nabla^2\vb{u}-\frac{N^2\rho_f}{\gamma}{\bf \hat{z}}+\vb{f}
\label{u-bouss}
\end{equation}

In a non-quiescent flow, the density field $\rho$ follows an
advection-diffusion equation with a diffusion constant $\kappa$. However, it is
useful to work, instead of the density field, with its surrogate, the buoyancy
field $b = \frac{\rho_f N}{\gamma}$. This buoyancy field has the added
advantage of being dimensionally the same as the velocity field. By
rewriting the advection-diffusion equation of $\rho$ in terms of $b$ and
rewriting the buoyancy term in Eq.~\eqref{u-bouss} in terms of $b$, we arrive
at the final set of coupled, equations which describe a (turbulent) stratified
flow

\begin{eqnarray}
	\pdv{\vb{u}}{t}+\vb{u}\cdot\nabla\vb{u} &=& -\nabla p+\nu\nabla^2\vb{u}-Nb \bf \hat{z}+\vb{f};\label{u}\\
	\pdv{b}{t}+\vb{u}\cdot\nabla b  &=& \kappa\nabla^2 b + N \vb{u}.\bf \hat{z};\label{b}
\end{eqnarray}
along with the incompressibility constraint
\begin{equation}
    \nabla\cdot\vb{u} = 0;\label{incomp}
\end{equation}

The external force ${\bf f}$ drives the system to a statistical steady, homogeneous but anisotropic turbulent state for a suitable choice of parameters. As discussed earlier, the anisotropy along the vertical direction $\bf{\hat{z}}$ leads to distinct statistical behavior when measured in planes along the $Z$-axis to those measured on the $XY$ plane perpendicular to $\bf{\hat{z}}$. To differentiate between these two directions, it is convenient to express quantities using the superscripts $\perp$ for components measured along the vertical direction and $\parallel$ for the component measured in the planes perpedicular to the $\bf{\hat{z}}$. For instance, the velocity field ${\bf u}$ can be decomposed in to components ${\bf u}^{\perp}$ and ${\bf u}^{\parallel}$. 

The sensitivity of the flow to stratification is determined by the competition between the buoyancy and inertial forces in Eq.~\eqref{u}. Thence, the degree of stratification is characterised by the non-dimensional Froude number $Fr =  \frac{u_{\rm rms}}{L_{\rm int}N}$ obtained from a simple ratio of inertial-to-buoyancy forces; here $u_{\rm rms}$ is the root-mean-square velocity, and $L_{\rm int}$ is the integral (large) length scale of the flow. The integral length scale is most conveniently expressed in terms of the energy spectrum $E(k)$ of turbulence through $L_{\rm int} = \frac{\int_0^\infty k^{-1}E(k)dk}{\int_0^\infty E(k)dk}$.

Just at the Froude number gives us a measure of the strength of stratification, the degree of turbulence is determined by the non-dimensional Reynolds number which is a measure of the ratio of the inertial to dissipative forces in Eq.~\eqref{u}. Typically, this measure is the Taylor-scale-based Reynolds number $Re_\lambda = u^2_{\rm rms}\sqrt{15/\epsilon_k\nu}$  where $\epsilon_k$ is the mean kinetic energy dissipation rate of the flow. Finally, the contrast between the fluid viscosity and density diffusivity is set by the Prandtl number $Pr = \nu/\kappa$. Strongly stratified turbulence implies that the  buoyancy Reynolds number $Re_b \sim \left (Re_\lambda Fr \right )^2 > 1$ and $Fr \ll 1$ along with $Re_\lambda \gg 1$.   

Finally, the  Brunt-V\"ais\"al\"a  frequency $N = \sqrt{\frac{g\gamma}{\rho_0}}$ has dimensions of inverse time.  In order to make comparisons between our simulations and observations meaningful, we define a non-dimensional form of this frequency ${\tilde N} \equiv \tau_\eta N$, where $\tau_\eta \equiv \sqrt{\frac{\nu}{\epsilon}}$ is the characteristic short (Kolmogorov) time-scale of the underlying flow.

To have a sense of these numbers in what follows, it is useful to have an estimate of the key parameters defining strongly stratified turbulence in, for example, an ocean. Indeed, it is well known that there exists huge variations in the typical velocity, stratification and dissipation rates across the ocean and at different depths. Nevertheless, in a typical ocean, it is known that the $\mathcal{O}(10^{-3}{\rm sec}) \lesssim N \lesssim \mathcal{O}(10^{-1}{\rm sec})$. Given that $\tau_{\eta} \sim 1 {\rm sec}$ (since for sea water $\nu = 10^{-6}{\rm m^2/sec}$ and $\epsilon_k \approx  10^{-6}{\rm m^2/sec^3}$), the normalised frequency also varies as $\mathcal{O}(10^{-3}) \lesssim {\tilde N} \lesssim \mathcal{O}(10^{-1})$. Observations suggest that  $0.1{\rm m/sec} \lesssim u_{\rm rms} \lesssim 2.0{\rm m/sec}$ and $\mathcal{O}(10^4) {\rm m} \lesssim L_{\rm int} \lesssim \mathcal{O}(10^5) {\rm m}$; hence the Froude numbers range as $\mathcal{O}(10^{-4}) \lesssim Fr \lesssim \mathcal{O}(10^{-2})$ and $Re_\lambda \sim \mathcal{O}(10^3)$. Finally, the Prandtl number can vary from $Pr \sim 7$ (thermal stratification) to as high as 700 (salt stratification). In the fully developed turbulent regime, it is reasonable to assume $Pr \sim 1$.

The equations governing stratified turbulence are solved by using direct
numerical simulations (DNSs) of Eqs.~\eqref{u}-\eqref{incomp}. The simulations
are performed using a pseudo-spectral method, on a triply periodic cube of side
length 2$\pi$, which is discretized into $512^3$ collocation points. We use a
second-order Runge-Kutta method for time-marching. The flow is driven to a
non-equilibrium, statistically stationary through a constant energy injection
force ${\bf f}$ limited to the first two wavenumber shells ($k \leq 2$), and
applied only to the horizontal scales.  A constant kinematic viscosity of $\nu
= 5 \times 10^{-3}$ and the constant energy injection $\epsilon = 0.9$ ensures,
in the steady state, the flow achieves a Taylor-scale Reynolds number
$Re_\lambda \approx 140$. We conduct simulations for different normalized
Brunt-V\"ais\"al\"a frequerncies, ${\tilde N} =$ 0.28, 0.35, 0.42, 0.49 and
0.56. These correspond to Froude numbers are $Fr =$ 0.238, 0.182, 0.158, 0.143
and 0.126. The Prandtl number is set to unity across all simulations. 

It is important to emphasize that the Froude numbers and (normalised)
Brunt-V\"ais\"al\"a  frequencies in our simulations falls well within the range
observed in oceanic environments. Nevertheless, the range of values is
constrained by the computational resources currently available to us. It is
easy to check, however, that all our simulations are within the strongly
stratified regime with $Re_b \gg 1$, $Fr \ll 1$ and $Re_\lambda \gg 1$. 

The density stratification and hence the emergent buoyancy force lead to a vertical confinement of fluid elements. The restraint on the motion 
of vertical mixing of fluid elements are accentuated with increasing Brunt-V\"ais\"al\"a  frequencies. This confinement is most easily seen 
when tracking fluid elements through Lagrangian tracers seeded in the flow, once a statistically steady, non-equilibrium state is reached. 
Such a point-like, massless tracer particle, characterised by its instantaneous position ${\bf x}$ and velocity ${\bf v}$ vectors, follows 
the Lagrangian dynamics
\begin{equation}
	\frac{d{\bf x}}{dt} = {\bf v}; \quad \quad {\bf v} = {\bf u}({\bf x},t).
	\label{lag}
\end{equation}
In our simulations, to obtain reliable statistics, we seed $10^5$ Lagrangian particles and numerically solve for the trajectories defined above by using 
a second-order Runge-Kutta scheme.  The fluid velocity ${\bf u}({\bf x},t)$, at the typically off-grid Lagrangian position ${\bf x}$ is 
obtained through a trilinear interpolation from the regular, periodic Cartesian grid on which the Euler velocity (and density) field equations are solved. 

\begin{figure}[t]
\includegraphics[width=1.0\linewidth]{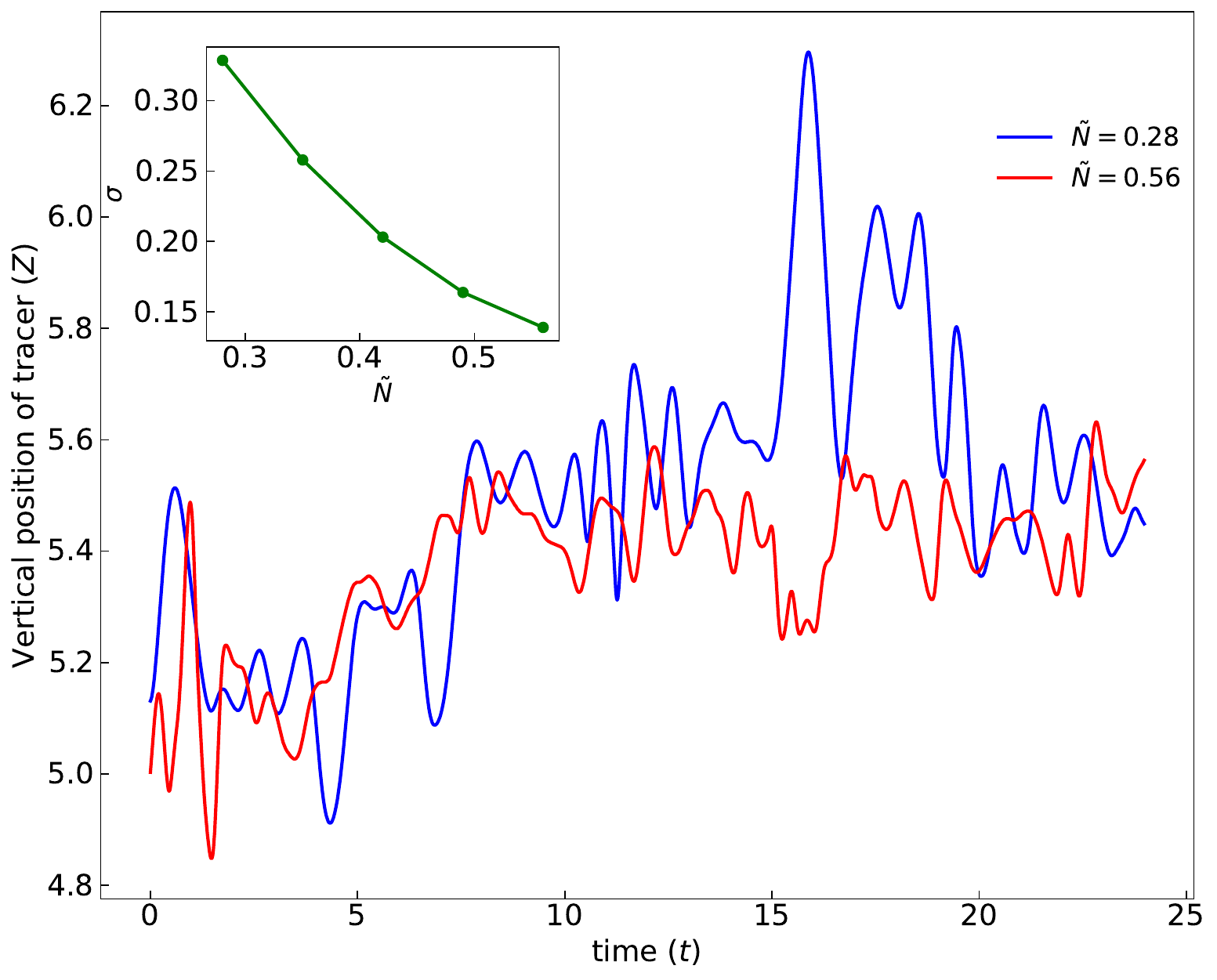}
	\caption{Time evolution of vertical position ($Z(t)$) for a tracer at extreme normalized Brunt-V\"ais\"al\"a frequencies $\tilde{N}=0.28$ and  $\tilde{N}=0.56$. Increasing $\tilde{N}$ suppresses the fluctuations and tend to confine the tracers vertically. The inset quantifies this effect, showing the  standard deviations of the vertical fluctuations decreasing with $\tilde{N}$ }
	\label{fig:traj}
\end{figure}

\begin{figure}[t]
	\includegraphics[width=1.0\linewidth]{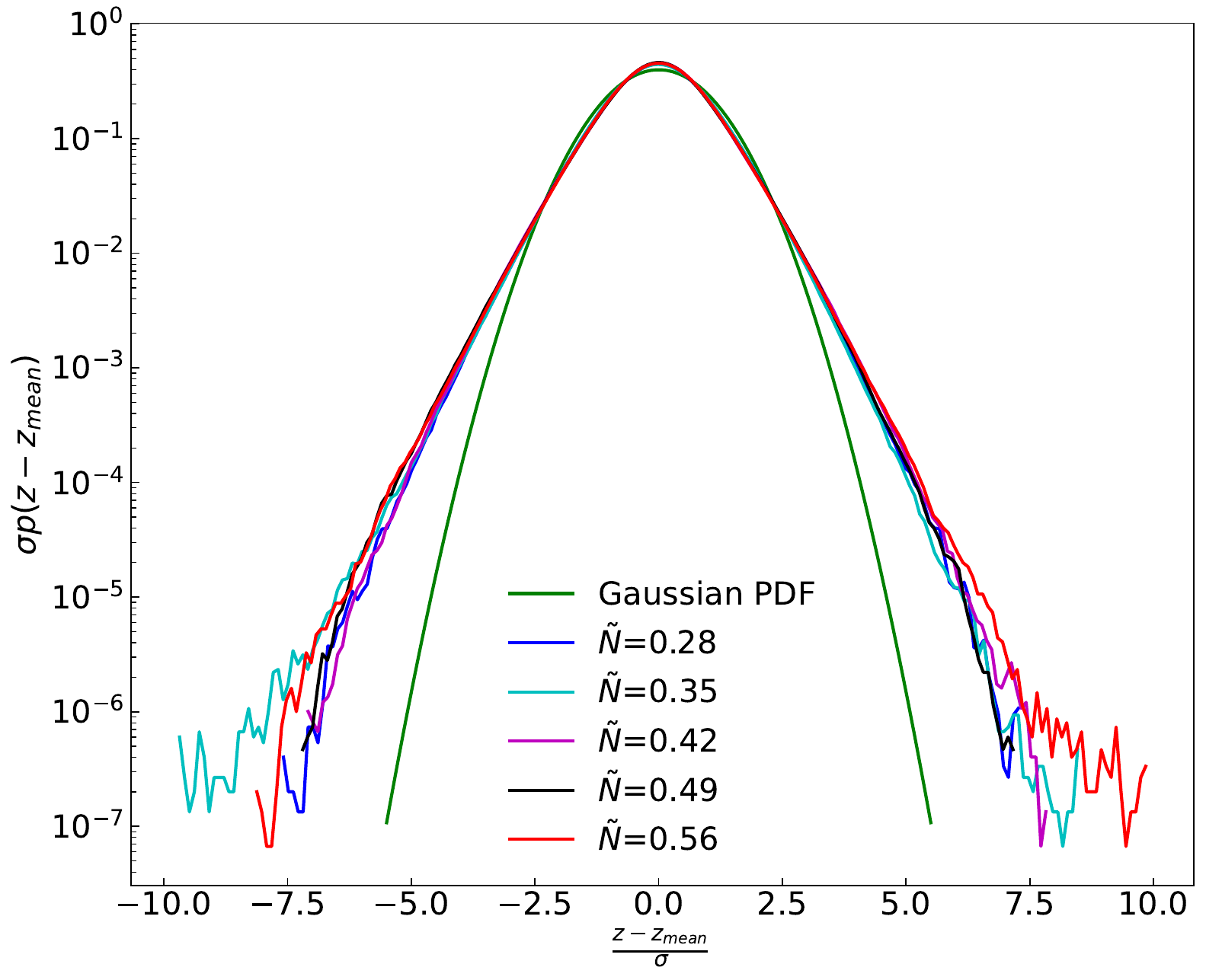}
	\caption{Probability distribution functions (PDFs) of normalized vertical displacements ($\frac{z-z_{mean}}{\sigma}$) for different stratification strengths $\tilde{N}$. The data collapse onto a universal curve, highlighting the scale-independence in strongly stratified turbulence. Deviations from the Gaussian distribution (green curve) reveal non-Gaussian tails, suggesting intermittent vertical mixing.}
	\label{fig:pdf}
\end{figure}

In Fig.~\ref{fig:traj}, we show representative plots of the vertical position
$Z$ of a randomly selected tracer trajectory from simulations with the smallest
$\tilde{N}=0.28$  and largest $\tilde{N}=0.56$ (normalized) Brunt-V\"ais\"al\"a
frequencies. The visual evidence of stronger confinement and subdued
fluctuations with increasing $\tilde{N}$ is of course suggestive but hardly
conclusive. To see if this is indeed the case, we calculate, for a given
$\tilde{N}$ and particle, the mean vertical position $\bar{z}$ and thence a
measure of the fluctuations through the standard deviation $\sigma = \langle
\left [z - \bar{z} \right ]^2 \rangle$, where the angular brackets $\langle
\cdot \cdot \cdot \rangle$ denotes an average over all $10^5$ trajectories. 

In the upper inset of Fig.~\ref{fig:traj}, we plot $\sigma$ vs $\tilde{N}$ and,
as anticipated, we find increasing confinement as the flow becomes more
stratified with increasing $\tilde{N}$. Using this, we compute the probability density functions of the
normalised vertical displacements $\tilde{z} \equiv \frac{z -
\bar{z}}{\sigma}$ and are shown in Fig. ~\ref{fig:pdf}. The PDFs exhibit distinct non-Gaussian features and collapse onto a common curve. 
For comparison, we also plot a Gaussian distribution, highlighting the deviations from Gaussianity, particularly in the tails of computed PDFs. 
The collapse for different $\tilde{N}$ also highlights the underlying statistical structure of the vertical confinement in the strongly stratified turbulence regime.   

\begin{figure}[hbt]
	\includegraphics[width=1.0\linewidth]{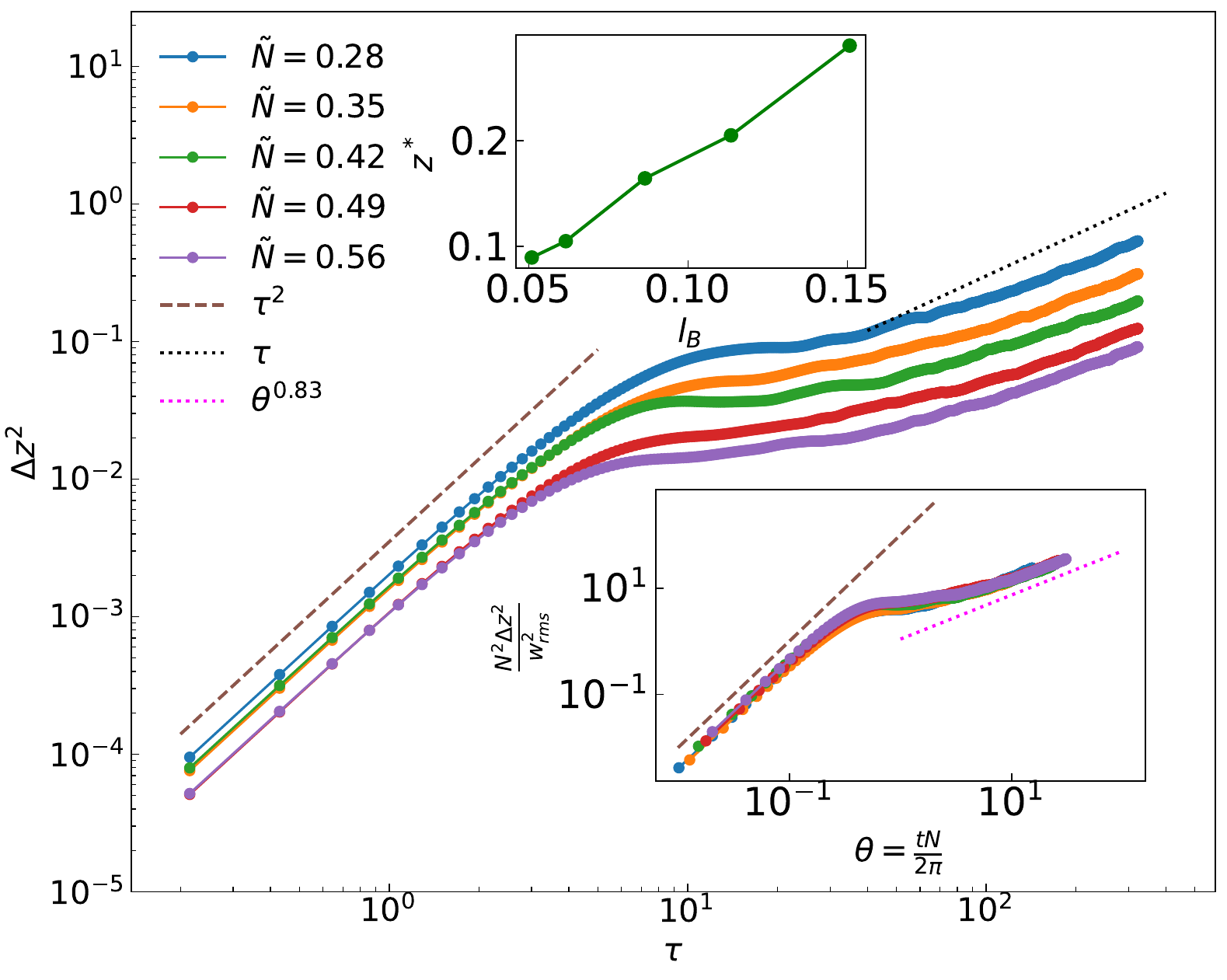}
	\caption{Vertical mean square displacement (MSD) of tracer particles as a function of the rescaled time ($\tau=\frac{t}{\tau_\eta}$). The main panel shows the ballistic regime ($\Delta z^2 \sim \tau^2$) at short times, transitioning to a sub-diffusive regime at longer times. The upper inset presents the estimated transition length scale($z^*$) against the buoyancy scale ($l_B$). The upper inset displays the MSD curves rescaled by $\tau_B$ and $l_B^2$, revealing a universal sub-diffusive scaling regime, $\theta^{\xi}$ with $\xi=0.83$.}
	\label{fig:MSD}
\end{figure}

How does this effect of vertical confinement show up in the (vertical) mean
square displacement (MSD) of trajectories?  We define the $\Delta z^2 =
\ang{\vert\vert {z}(t) - z(0)\vert\vert^2}$, where the angular brackets
$\ang{\cdot}$ denote an ensemble averaging over all trajectories and
$\vert\vert\cdot\vert\vert$ the Eulerian norm. We expect $\Delta z^2$ to show a
ballistic scaling $\tau^2$ (where we use non-dimensional time $\tau =
t/\tau_\eta$) for short times and up to separations $\Delta z^2 \sim z_*^2$.
This ballistic regime is clearly seen in Fig.~\ref{fig:MSD} which terminates at
a Brunt-V\"ais\"al\"a frequency dependent length scale $z_*$.  From physical
grounds it is reasonable to assume that this ballistic regime persists up to
length scales $z_*$ which are comparable to the intrinsic buoyancy scale
$\ell_B$ of the flow.  The buoyancy length scale is estimated, dimensionally,
from the root-mean-square vertical fluid velocity $u_z^{\rm rms}$ via $\ell_B
\sim \frac{u_z^{\rm rms}}{N}$. In order to test this hypothesis, we show a plot
of $z_*$, estimated from plots of $\Delta z^2$, vs $\ell_B$ in the upper inset
of Fig.~\ref{fig:MSD}. The reasonable compelling linear relation between the
two confirms our conjecture $z_* \approx \ell_B$.  Beyond $z_*$, and at longer
times, the MSD shows distinct sub-diffusive behaviour  as is easily 
seen when compared to the $\Delta z^2 \sim t$ line drawn in Fig.~\ref{fig:MSD} as a 
guide to the eye. 

Is it reasonable to conjecture that the transition from the ballistic to the
sub-diffusive regime is set by the timescale $\tau_b = 2\pi/N$ associated with the
typical time period of the \textit{oscillatory} fluid elements in the vertical
direction. This would suggest that the MSD plots shown in the main panel of
Fig.~\ref{fig:MSD} ought to collapse on rescaling time with $\tau_b$ (and  $\Delta
z^2$ by $l_B^2$). In the lower inset of Fig.~\ref{fig:MSD} we show plots of the
MSD (as in the main panel) but on such rescaling.  Furthermore, such rescaled
plot suggest a clear sub-diffusive regime $\Delta z^2 \sim \theta^\xi$, with $\xi
\approx 0.83$ for all Brunt-V\"ais\"al\"a frequencies. Here, we have denoted the rescaled time as $\theta=\frac{t N}{2\pi}$. (Our calculations for
the MSD on the horizontal planes are  in agreement to those reported earlier by
van Aartrijk, Clercx, and Winters~\cite{van2008single}.) 

\begin{figure*}[hbt]
	\centering
	\begin{subfigure}{0.49\linewidth}
		\centering
		\includegraphics[width=\linewidth]{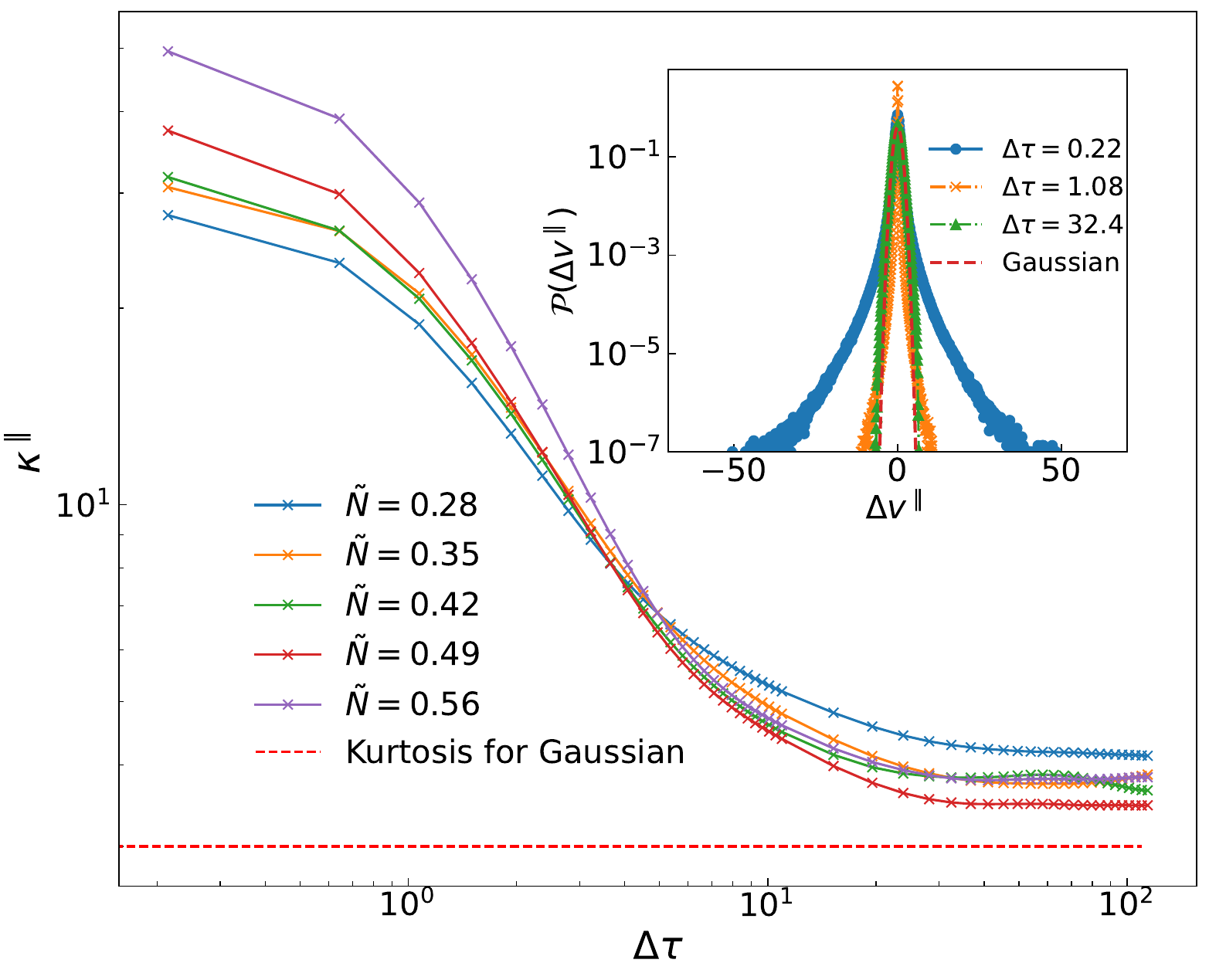}
		\caption{}
	\end{subfigure}
	\hfill
	\begin{subfigure}{0.49\linewidth}
		\centering
		\includegraphics[width=\linewidth]{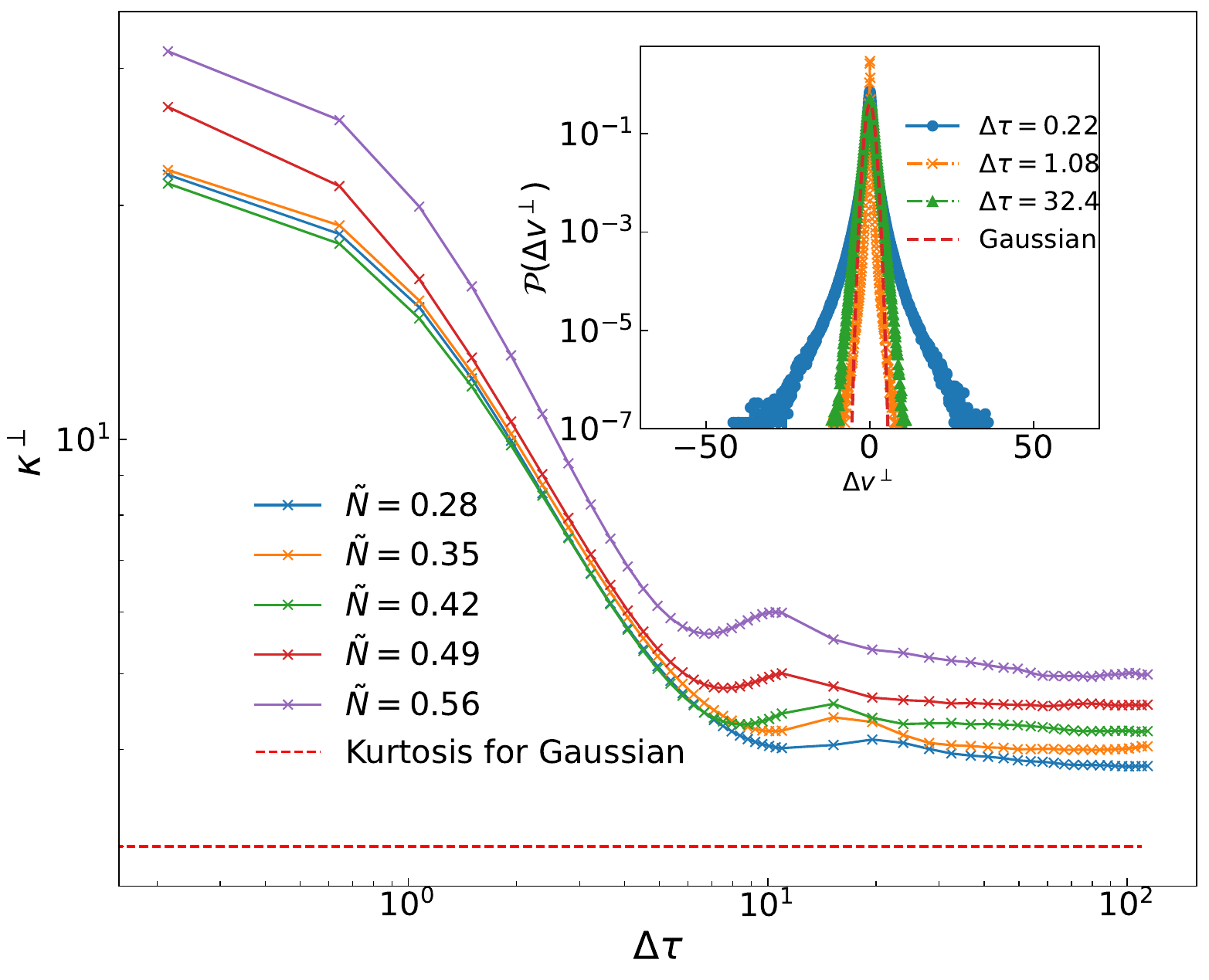}
		\caption{}
	\end{subfigure}
	\caption{Velocity increment statistics revealing intermittency in stratified turbulence. (a) Kurtosis $\kappa^\parallel$ of horizontal velocity increments ($\Delta v^{\parallel}$) and (b) $\kappa^\perp$ of vertical  increments ($\Delta v^{\perp}$) versus normalized time increments ($\Delta \tau$); In the inset of each main panel, the corresponding PDFs ($\mathcal{P}^\parallel$ and $\mathcal{P}^\perp$)  at three $\Delta \tau$ values are shown for largest $\tilde{N}(=0.56)$. Widening non-Gaussian tails(insets) and elevated Kurtosis values at small $\Delta \tau$ demonstrate strong intermittency.}
	\label{fig:increment}
\end{figure*}

We now turn our attention to the Lagrangian velocity statistics to uncover
how intermittent such flows are and their dependence on the Brunt-V\"ais\"al\"a
frequency. We begin by examing the Lagrangian velocity increments separately
for the vertical and horizontal components of the velocity field. These
increments are defined, respectively, as $\Delta v ^{\perp} \equiv v_z(t +
\Delta t) - v_z(t)$ and $\Delta v ^{\parallel} \equiv \left (v_x(t + \Delta t)
- v_x(t)\right ) + \left (v_y(t + \Delta t) - v_y(t)\right )$.  In the insets
of Fig~\ref{fig:increment} we show the corresponding probability density
functions (a) $\mathcal{P}^\perp$ and (b) $\mathcal{P}^\parallel$ for three
different (normalised) time increments $\Delta \tau = \frac{\Delta
t}{\tau_\eta}$. While for large $\Delta \tau$ we expect the distributions to
approach a Gaussian, the nature of these distributions for small $\Delta \tau$
holds the key to the nature of Lagrangian intermittency in strongly stratified
turbulence. The insets of Figures~\ref{fig:increment}(a) and (b) show clear non
Gaussian, and ever widening tails of the distributions as $\Delta \tau \to 0$.
Such non Gaussian tails are typically signatures of Lagrangian intermittency 
which are quantified through a measure of the kurtosis for both the 
vertical $\kappa^{\perp} = \frac{\langle \left ( \Delta v ^{\perp}\right )^4 \rangle}{\langle \left ( \Delta v ^{\perp}\right )^2 \rangle^2}$ 
and in-plane 
$\kappa^{\parallel} = \frac{\langle \left ( \Delta v ^{\parallel}\right )^4 \rangle}{\langle \left ( \Delta v ^{\parallel}\right )^2 \rangle^2}$
components. 
In the main panels of Fig.~\ref{fig:increment} we show plots of (a) $\kappa^{\perp}$ and (b) $\kappa^{\parallel}$ vs the normalised 
time increment $\Delta \tau$. Clearly as $\Delta \tau \to 0$, and taking the value of kurtosis as a surrogate for the degree 
of intermittency, the in-plane Lagrangian motion appears to be more intermittent than its vertical component as a possible consequence 
of the confinement discussed before with $\kappa^{\perp}$ and $\kappa^{\parallel}$ increasing with $\tilde{N}$ as $\Delta \tau \to 0$. 

Finally, we make use of the measurements reported in Fig.~\ref{fig:increment} to calculate the $p$-order, Lagrangian structure functions $S_p^{\parallel} \equiv \langle \left ( \Delta v ^{\parallel}\right )^p \rangle$ and $S_p^{\perp} \equiv \langle \left ( \Delta v ^{\perp}\right )^p \rangle$.   The quenched  motion in the vertical direction suggests that the horizontal dynamics dominate the flow leading to a power law scaling of the form $S_p^{\parallel} \sim \Delta \tau^{\zeta_p^\parallel}$. This scaling  holds for $1 \lesssim \Delta \tau \lesssim 40$ corresponding to the inertial range of the horizontal dynamics. For smaller $\Delta \tau$, the structure functions are expected to follow  a simple scaling derived from the Taylor expansion. For $\Delta \tau$ beyond the inertial range,  the structure functions ought to saturate.

\begin{figure}[t] 
	\includegraphics[width=\linewidth]{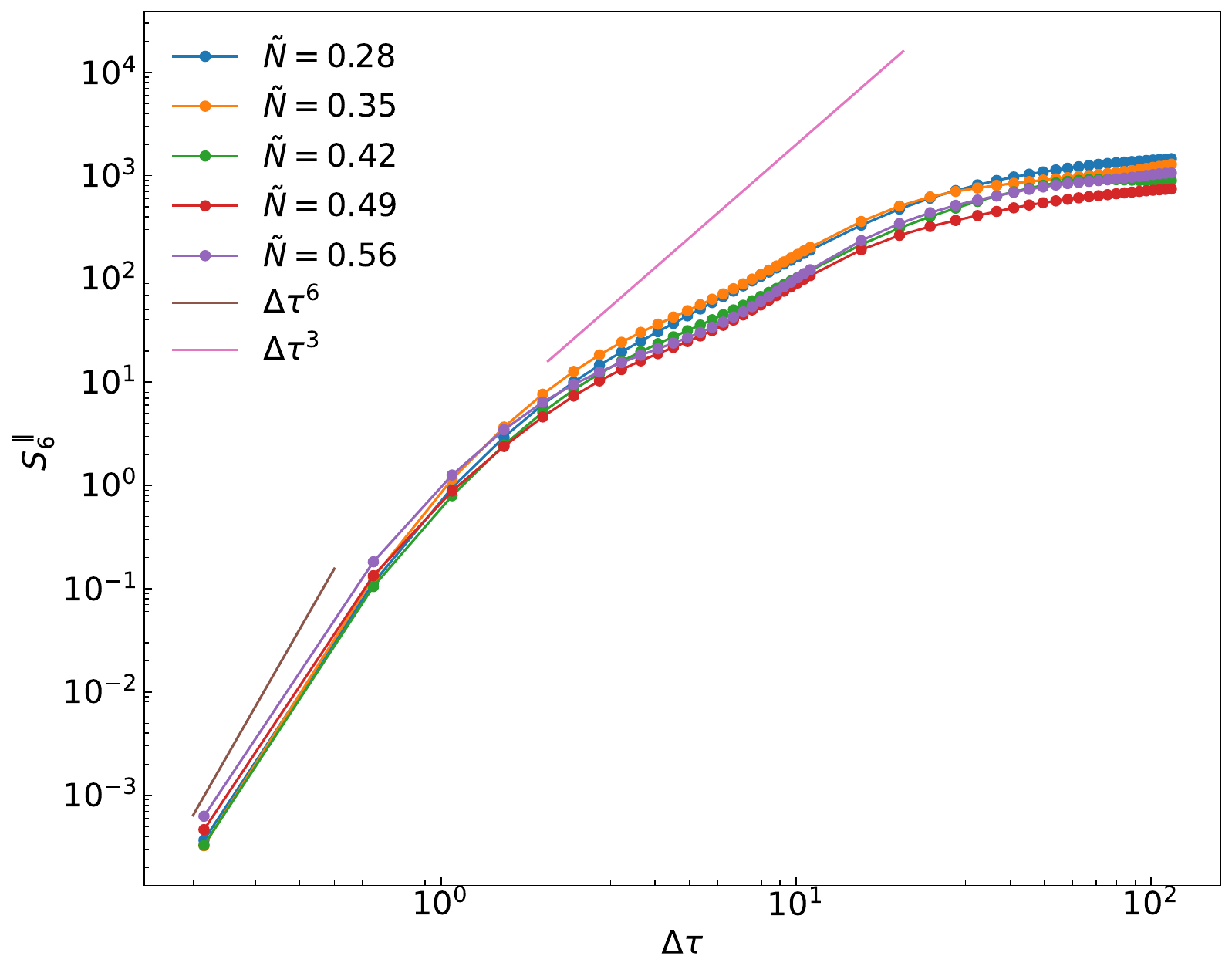}
	\caption{Scaling behavior of the sixth order horizontal Lagrangian velocity structure functions ($S_6^{\parallel}$). The figure also includes reference lines indicating the expected scaling in the limit of $\Delta \tau \rightarrow 0$ ($\zeta_6^{\parallel}=6$) and the inertial range scaling in the absence of intermittency ($\zeta_p^{\parallel}=3$).  The scaling from simulations show large deviation due to intermittency.}
	\label{fig:sixthstruct}
\end{figure}

From the dimensional analysis, we expect that horizontal structure functions should scale as $S_p^{\parallel} \sim (\epsilon \Delta \tau)^{p/2}$  in the inertial range in the absence of intermittency,  which correspond to scaling exponents $\zeta_p^{\parallel}=p/2$.  To test this scaling behavior, we plot the sixth order structure function $S_6^{\parallel}$ against $\Delta \tau$ in Fig.~\ref{fig:sixthstruct}.  This figure also include two reference plots:  the expected scaling in the limit of $\Delta \tau \rightarrow 0$ ($\zeta_6^{\parallel}=6$),  and the inertial range scaling ($\zeta_6^{\parallel}=3$) in the absence of intermittency. While the small-$\Delta \tau$ data agree well with the $\zeta_6^{\parallel}=6$ scaling, significant deviations from $\zeta_6^{\parallel}=3$ emerge in the inertial range, demonstrating strong intermittency at this order. Similar behavior occurs for lower orders (not shown here), with decreasing intermittency effects as the order ($p$) reduces.  

To further characterize the intermittency of the predominantly horizontal motion, we analyze the scaling exponent ratios $\zeta_p^{\parallel}/ \zeta_2^{\parallel}$ using the extended self similarity (ESS) \cite{biferale2006lagrangian,RayESS} combined with local slope analysis.  The local slopes are defined as:  $\frac{\zeta_p^{\parallel}}{\zeta_2^{\parallel}}(\Delta \tau) = \frac{d\log(S_p^\parallel(\Delta \tau))}{d\log(S_2^\parallel(\Delta \tau))}$. Examining these slopes across varying time separations ($\Delta \tau$) provides insight into the accuracy and the scaling behavior of the exponents. 

\begin{figure*}[t]
	\centering
	\begin{minipage}{0.48\linewidth}
		\centering
		\vspace{-0.1in}
		\includegraphics[width=\linewidth]{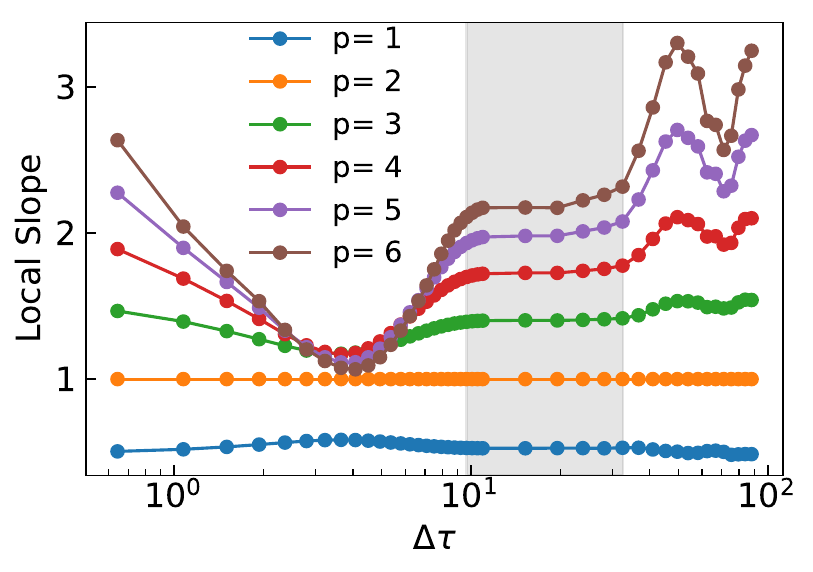}
		\caption{Local slopes of ratios of the scaling exponents for horizontal velocity structure functions for$\tilde{N}=0.56$. At small $\Delta \tau$,  slopes approach $p/2$(Taylor expansion limit). The emergence of a clear scaling range (in gray) at larger $\Delta \tau$ reveals power law behavior.}
		\label{fig:LocalSlope}
	\end{minipage}
	\hfill
	\begin{minipage}{0.48\linewidth}
		\centering
		\includegraphics[width=\linewidth]{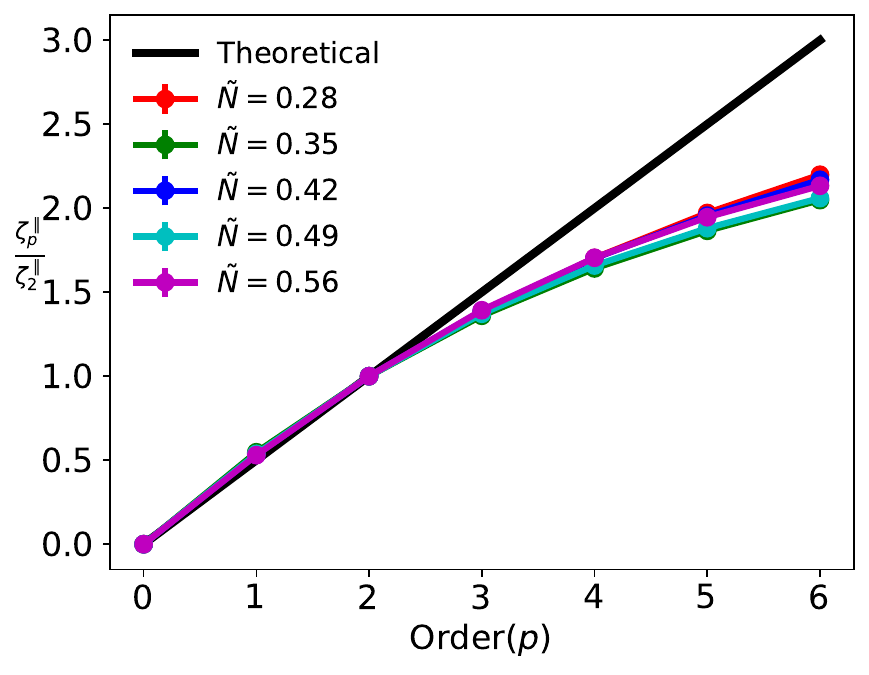}
		\caption{Ratio of scaling exponents, $\frac{\zeta_p^{\parallel}}{\zeta_2^{\parallel}}$ versus order $p$ for different $\tilde{N}$. Significant deviation from the theoretical non-intermittent scaling (black line) at higher orders confirms strong intermittency in horizontal Lagrangian velocities.}
		\label{fig:LagESS}
	\end{minipage}
\end{figure*}

Fig.~\ref{fig:LocalSlope} shows the local slopes for the largest $\tilde{N}(=0.56)$ in our simulations.  As time separations $\Delta \tau \rightarrow 0$, the local slopes approach $p/2$ consistent with Taylor-series expansion of the structure functions. As $\Delta \tau$ increases, the local slopes for $p>2$ initially decrease -- reflecting the influence of the vortical structures trapping the tracers \cite{benzi2023lectures} -- before rising again due to inertial effects, a behavior also observed in homogeneous isotropic turbulence.  A clear scaling range (highlighted in gray in Fig.~\ref{fig:LocalSlope}) emerges at larger $\Delta \tau$, indicating  power law scaling (with intermittency) in the exponent ratios. From this range, we extract  $\zeta_p^{\parallel}/\zeta_2^{\parallel}$ and plot it against the order $p$ in Fig.~\ref{fig:LagESS}. The theoretical non-intermittent scaling (black line) is significantly deviated at higher orders, confirming intermittency in the horizontal Lagrangian velocity structure functions. These results suggest a multifractal nature of the predominantly horizontal flow, motivating future work on detailed multifractal analysis in stratified turbulence.

\begin{figure}[t]
	\includegraphics[width=1.0\linewidth]{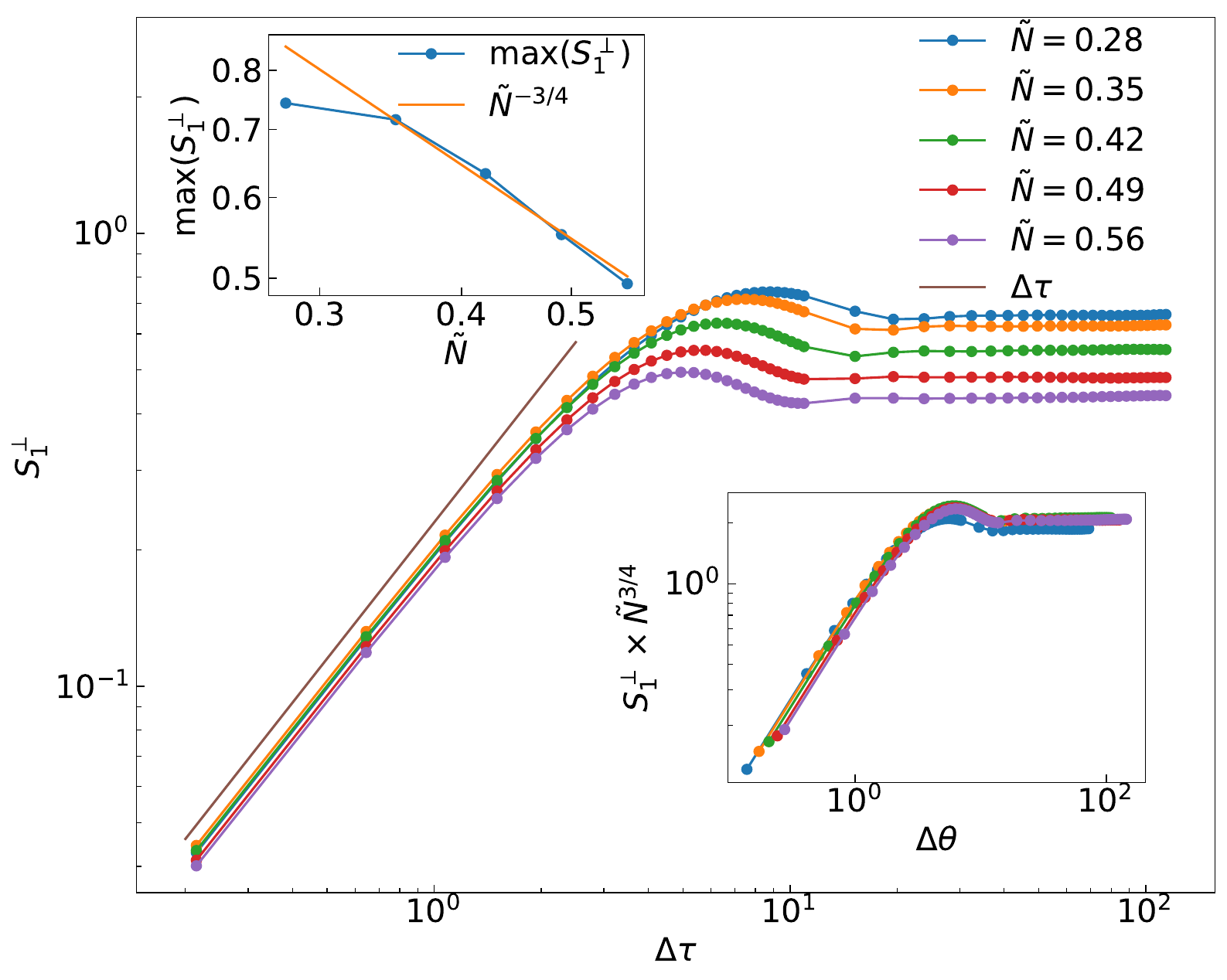}
	\caption{Vertical Lagrangian velocity structure function $S_1^{\perp}$ as a function of time increment $\Delta \tau$, showing $\tilde{N}$ dependent peak values, while the lower inset demonstrates the collapse of rescaled $S_1^\perp$ for $\Delta \theta \gtrsim 1 $.}
	\label{fig:verticalsecstruct}
\end{figure}

In addition to the horizontal Lagrangian velocity structure functions, we also examine their vertical velocity structure functions $S_p^{\perp}$. Unlike the horizontal structure functions, which exhibit power law scaling in the inertial range, the vertical counterparts show a distinct dependence on $\tilde{N}$. In the main panel of Fig.~\ref{fig:verticalsecstruct}, we show $S_1^{\perp}$ plotted against $\Delta \tau$; this indeed suggests that the peaks in $S_1^{\perp}$ depend on $\tilde{N}$ with the peaks appearing at $\tilde{N}$ dependent $\Delta \tau$. To quantify this dependence, we plot the maximums of $S_1^\perp$ against $\tilde{N}$  in the upper inset of  Fig.~\ref{fig:verticalsecstruct}. This suggest  $\tilde{N}^{-3/4}$ scaling for $S_1^{\perp}$ in a range where buoyancy forces are important.  In lower inset of  Fig.~\ref{fig:verticalsecstruct}, we show  $S_1^{\perp}$  (rescaled by $N^{-3/4}$) against the time increments rescaled by buoyancy time scale ($\Delta \theta$)  which indeed suggests a collapse of all the curves especially for $\Delta \theta \gtrsim 1$ .   The peaks too appear approximately the same $\Delta \theta$. The behavior of remaining Lagrangian vertical structure functions (not shown) follows a similar trend, with the scaling now given by $S_p^{\perp} \propto\tilde{N}^ {-\frac{3p}{4}}$ for the $\Delta \theta \gtrsim 1$. This indicates that the vertical motion, while constrained by buoyancy forces, still exhibits a systematic dependence on $\tilde{N}$.

To summarize, our findings demonstrate that increasing stratification strength enhances vertical confinement of tracers, leading to predominantly horizontal motion. This underscores the need for a detailed investigation of Lagrangian velocity increments in the horizontal direction. Our analysis reveals that intermittency in stratified turbulence exhibits distinct characteristics compared to homogeneous isotropic turbulence. Specifically, in the strongly stratified regime considered here, horizontal Lagrangian velocity increments display extended self-similarity, with pronounced intermittency at higher orders. While computationally demanding, extending this study to higher Reynolds numbers $Re_\lambda$ and smaller $Fr$ values would provide further insights into the nature of intermittency in strongly stratified turbulence. 

The present results contribute to the broader understanding of intermittency in stratified turbulence and its implications for particle transport, including heavy, active, and neutrally buoyant particles. Further investigation into these effects would enhance our comprehension of mixing and transport processes in geophysical and environmental flows. Additionally, a systematic study of Lagrangian tracer statistics, incorporating the effects of rotation and double diffusion, presents a promising avenue for future research. Complementary to this, examining intermittency in Eulerian statistics of horizontal velocity increments would establish a more comprehensive perspective on strongly stratified turbulence. 

Building on these results, we will extend our work by analyzing the intermittency's role on the orientation statistics of spheroids in stratified turbulence motivated by recent findings \cite{varanasi2022rotation} on sedimenting spheroids in stratified fluids. Prior work on sedimenting spheroids in homogeneous isotropic turbulence has shown that the gravity induced inertial torque leads to non-gaussian orientation distributions localized about the broadside-on orientation \cite{anand2020orientation}.  Stratification is expected to modify this behavior due to competing inertial and stratification torques that oppose each other in aligning spheroid axes \cite{varanasi2022rotation}. The underlying (stratified) turbulent torque can further modify this alignment. We will quantify these combined effects on the orientation statistics. In addition to the above, we will also investigate the elastic collisions between heavy (spherical) particles, which have been studied previously \cite{bec2013sticky} in three dimensional random incompressible flow and we expect these collisions to become more prevalent in stratified turbulence. 

\begin{acknowledgments}
	The author acknowledges Prof. Samriddhi Sankar Ray, ICTS-TIFR for his invaluable guidance, insightful discussions, and significant contributions to the development and writing of this manuscript. His feedback and criticism greatly improved the clarity of this work. The author also would like to thank Prof. Rama Govindarajan, ICTS-TIFR for many useful discussions. The simulations were performed on the ICTS clusters \emph{Tetris} and \emph{Contra}. This research was supported in part by the	International Centre for Theoretical Sciences (ICTS) for participating in the programs ---  \textit{Field Theory and Turbulence} (code:ICTS/ftt2023/12) and \textit{Turbulence: Problems at the	Interface of Mathematics and Physics} (code: ICTS/TPIMP2020/12). We acknowledge the support of the DAE, Govt. of India, under project no. 12-R\&D-TFR-5.10-1100 and project no. RTI4001.
\end{acknowledgments}

\bibliographystyle{unsrt}
\bibliography{LagrangianIntermittency}
\end{document}